\documentclass{emulateapj}

\def\gsim{\mathrel{\rlap{\lower 4pt \hbox{\hskip 1pt $\sim$}}\raise 1pt
\hbox {$>$}}}
\def\lsim{\mathrel{\rlap{\lower 4pt \hbox{\hskip 1pt $\sim$}}\raise 1pt
\hbox {$<$}}}

\begin{document}

\title{Nebular Spectra and Explosion Asymmetry of \\ 
Type Ia Supernovae}

\author{
K.~Maeda\altaffilmark{1}, 
S.~Taubenberger\altaffilmark{2}, 
J. ~Sollerman\altaffilmark{3,4}, 
P.A.~Mazzali\altaffilmark{2,5,6}, \\
G.~Leloudas\altaffilmark{4}, 
K.~Nomoto\altaffilmark{1}, 
K.~Motohara\altaffilmark{7}
}

\altaffiltext{1}{Institute for the Physics and Mathematics of the 
Universe (IPMU), University of Tokyo, 
5-1-5 Kashiwanoha, Kashiwa, Chiba 277-8568, Japan; 
keiichi.maeda@ipmu.jp .}
\altaffiltext{2}{Max-Planck-Institut f\"ur Astrophysik, 
Karl-Schwarzschild-Stra{\ss}e 1, 85741 Garching, Germany.}
\altaffiltext{3}{The Oskar Klein Centre, Department of Astronomy, 
Stockholm University, AlbaNova, 10691 Stockholm, Sweden.}
\altaffiltext{4}{Dark Cosmology Centre, Niels Bohr Institute, Copenhagen University, 
Juliane Maries Vej 30, 2100 Copenhagen, Denmark}
\altaffiltext{5}{Scuola Normale Superiore, Piazza Cavalieri 7, 56127 Pisa, Italy}
\altaffiltext{6}{INAF - Oss. Astron. Padova, vicolo dell'Osservatorio, 5, 35122 Padova, Italy}
\altaffiltext{7}{Institute of Astronomy, Graduate School of Science, University of Tokyo, Mitaka, Tokyo 181-0015, Japan}

\begin{abstract}
The spectral signatures of asymmetry in Type Ia Supernova (SN Ia) explosions 
are investigated, using a sample of late-time nebular spectra. 
First, a kinematical model is constructed for SN Ia 2003hv, 
which can account for the main features in 
its optical, Near-Infrared (NIR), and Mid-Infrared (Mid-IR) 
late-time spectra. It is found that an asymmetric off-center model can 
explain the observed characteristics of SN 2003hv.
This model includes a relatively high density, Fe-rich region which displays a large velocity off-set, and a relatively low density, extended $^{56}$Ni-rich region which is more spherically distributed.
The high density region consists of the inner stable Fe-Ni region and outer $^{56}$Ni-rich region. 
Such a distribution may be the result of a delayed-detonation explosion, 
in which the first deflagration produces the global asymmetry 
in the innermost ejecta, 
while the subsequent detonation can lead to the bulk spherical symmetry. 
This configuration, 
if viewed from the direction of the off-set, can 
consistently explain the blueshift in some of the emission lines and 
virtually no observed shift in other lines in SN 2003hv. 
For this model, we then explore the 
effects of different viewing angles and the implications for 
SNe Ia in general. 
The model predicts that a variation of the central wavelength, 
depending on the viewing angle, should be seen 
in some lines (e.g., [Ni~II]~$\lambda$7378), 
while the strongest lines (e.g., [Fe~III] blend at $\sim 4700$\AA) 
will not show this effect. By examining optical 
nebular spectra of 12 SNe~Ia, we have found that such a variation indeed exists. 
We suggest that the global asymmetry in the innermost ejecta, as likely 
imprint of the deflagration flame propagation, is a generic feature of SNe~Ia. 
It is also shown that various forbidden lines in the NIR and Mid-IR regimes
provide strong diagnostics to further constrain the explosion geometry 
and thus the explosion mechanism.
\end{abstract}

\keywords{radiative transfer -- 
nuclear reactions, nucleosynthesis, abundances --
supernovae: individual (SN 2003hv) -- 
supernovae: general 
}

\section{INTRODUCTION}

Type Ia supernovae (SNe~Ia) provide a powerful tool to investigate the 
cosmological parameters and the nature of the dark energy 
(Riess et al. 1998; Perlmutter et al. 1999). 
Thanks to the uniformity of their optical 
peak luminosity, once a phenomenological relation between 
the light curve shape and the peak luminosity is applied, 
SNe~Ia can be used as reliable cosmological distance indicators 
(Phillips 1993; Phillips et al. 1999). 

There is a general consensus that SNe~Ia 
are thermonuclear explosions of a carbon-oxygen white dwarf (WD) 
(e.g., Nomoto et al. 1994, 1997; Wheeler et al. 1995; 
Branch 1998; Hillebrandt \& Niemeyer 2000). 
A Chandrasekhar-mass WD has been favored as a progenitor 
for the majority of SNe~Ia 
(e.g., H\"oflich \& Khokhlov 1996; 
Nugent et al. 1997; Mazzali et al. 2007). 

The thermonuclear runaway starts with the ignition of deflagration bubbles 
(e.g., Nomoto et al. 1976, 1984). 
It has been suggested that the deflagration flame may turn into 
a detonation wave 
(delayed detonation model, or deflagration-detonation-transition model; 
Khokhlov 1991; Yamaoka et al. 1992; Woosley \& Weaver 1994; Iwamoto et al. 1999; 
R\"opke \& Niemeyer 2007b), 
although the details of the transition have not been clarified yet. 

\begin{deluxetable*}{llll}
 \tabletypesize{\scriptsize}
 \tablecaption{SN~Ia Optical Nebular Spectra Sample\tablenotemark{a}
 \label{tab:tab1}}
 \tablewidth{0pt}
 \tablehead{
   \colhead{SN}
 & \colhead{Phase (Days)}
 & \colhead{[Ni~II]\tablenotemark{b}}
 & \colhead{References} 
}
\startdata
1986G & +103, +257 & Yes & Cristiani et al. (1992) \\
1990N & +186, +227, +255, +280, +333 & Yes & G\'omez et al. (1996)  \\ 
1991bg & +117, +199 & No & Turatto et al. (1996) \\
1991T & +258, +284, +316 & No & G\'omez et al. (1996) \\
1992G & +106, +128 & No & G\'omez et al. (1996) \\
1994D & +106 & Yes & G\'omez et al. (1996) \\
1996X & +246 & No & Salvo et al. (2001) \\
1998aq & +211, +231, +241 & Yes & Branch et al. (2003) \\
1998bu & +249, +329 & Yes & Cappellaro et al. (2001) \\
2000cx & +125, +147, +360 & Yes & Li et al. (2001); Sollerman et al. (2004) \\
2001V & +106 & Yes & Matheson et al. (2008) \\
2001el & +398 & Yes & Mattila et al. (2005) \\
2002dj & +275 & Yes & Pignata et al. (2008) \\
2002er & +216 & No & Kotak et al. (2005) \\
2003cg & +385 & No & Elias-Rosa et al. (2006) \\
2003du & +138, +141, +209, +221, +272, +377 & Yes & Anupama et al. (2005); 
Stanishev et al. (2007) \\ 
2003hv & +110, +143, +320, +358, +398\tablenotemark{c} & Yes & 
Leloudas et al. (2009); Gerardy et al. (2007); Motohara et al. (2006) \\
2004dt & +152 & Yes &  Altavilla et al. (2007) \\
2004eo & +228 & No & Pastorello et al. (2007) \\
2005cf & +267 & No & Leonard (2007) \\
\enddata
\tablenotetext{a}{The data are compiled from the {\it SUSPECT} data base, 
except for SNe 2000cx, 2001el and 2003hv. For SN 2003hv, NIR data are also 
given in this Table.}
\tablenotetext{b}{Whether the emission line feature at the red end 
of [Fe~II]~$\lambda$7155, 
which we identify as [Ni~II]~$\lambda$7378 (\S 4 \& 5), is detected or not at least at one epoch.}
\tablenotetext{c}{The spectra at $358$ days and $398$ days are the Mid-IR and NIR ones, respectively. }
\end{deluxetable*}

The first deflagration phase may well proceed in a very asymmetric way 
(Niemeyer et al. 1996; 
Garc\'ia-Senz \& Bravo 2005; Livne et al. 2005; 
Jordan et al. 2008). 
The deflagration wave propagates under the work of the 
buoyancy force, and thus a small perturbation 
in the progenitor structure could result in a global asymmetry. 
Rotation and convection in the progenitor WD could provide 
the seed for this asymmetric deflagration propagation.
For example, there is a possibility that 
the convection in the progenitor WD is dominated by a dipole mode 
(Woosley et al. 2004), which likely results 
in highly off-axis ignition and propagation of the 
deflagration flame (R\"opke et al. 2007c; Kasen et al. 2009).

It is, however, observationally challenging to put constraints on the
geometry of the explosion.
Measurements of polarization suggest 
that a large global asymmetry does not exist in 
SNe~Ia (Wang et al. 1996), 
with only a few exceptions (Howell et al. 2001). 
However, the polarization 
probes mainly the outer regions 
of the expanding SN ejecta, 
at least with the existing telescopes and instruments. 
The signature of the possible asymmetry in the deflagration phase, 
however, can only be 
probed by looking deeper into the innermost regions. 
In this respect, late-phase ($\sim 1$ year past the explosion) 
spectroscopy can provide an important diagnostics. 
Following the homologous expansion, 
the SN ejecta become transparent to optical and 
longer wavelengths, and thus emission line profiles 
can be used to probe the distribution of 
elements that emit the light or that deposit the energy. 
This strategy has been applied to investigate 
the aspherical nature of core-collapse SNe from massive stars (e.g., 
Maeda et al. 2002; Chugai et al. 2005; Mazzali et al. 2005; Maeda et al. 2008; 
Modjaz et al. 2008; Taubenberger et al. 2009; but see also Milisavljevic et al. 2009). 

Probing the explosion geometry of SNe Ia using late-time 
spectroscopy is still a fairly young field. 
H\"oflich et al. (2004) presented Near-Infrared (NIR) spectra 
for SN Ia 2003du taken with the {\it Subaru} telescope. 
They discussed that a flat-topped profile 
of  the [Fe~II]~1.644~$\micron$ emission feature 
is likely a result of a hole in the distribution of $^{56}$Ni. 
Motohara et al. (2006) added two other Near-Infrared (NIR) 
spectra for SNe~Ia taken with the {\it Subaru} telescope 
(SNe 2003hv, and 2005W).  
The [Fe~II]~1.257~$\micron$ and 1.644~$\micron$ emission lines displayed
different profiles, and especially, those in SN~2003hv were blueshifted by 
$\sim 2,000 - 3,000$~km~s$^{-1}$. 
Motohara et al. (2006)
suggested that this can be interpreted as 
evidence of a global asymmetry in SN~2003hv. 
Gerardy et~al.~(2007) also found a similar blueshift, 
in the [Co~III]~$11.88~\micron$ line detected in their Mid-Infrared (Mid-IR) 
spectrum of SN 2003hv, taken with the {\it Spitzer Space Telescope}. 
On the other hand, the late-phase optical spectra, 
available for more than 20 normal SNe~Ia, 
all look very similar to each other,
and seem to show little evidence for any asymmetry.

Recently, Leloudas et al. (2009) reported on observations of SN~2003hv, 
including late-time optical spectra. 
Thus, it is now for the first time possible to examine a late-time spectrum of 
a SN~Ia all the way from the optical to the Mid-IR.
We want to investigate if the kinematical interpretation for 
the blueshifts in the NIR and Mid-IR forbidden lines of SN~2003hv 
is consistent with the optical spectrum. 

In this paper, we examine late-time spectra of SN 2003hv, 
using multi-dimensional radiation transfer calculations 
and a simple kinematic model, and arrive at a plausible explosion 
geometry of SN 2003hv. 
Using the structure that was successfully applied to SN~2003hv, 
we then investigate signatures of the 
asymmetry and the expected diversity in the late-phase spectra, 
resulting from various viewing orientations. 
We also compile published late-time optical spectra of SNe~Ia, and investigate to what extent the expected diversity exists in the data.
In doing this, we identify a possible signature of the explosion asymmetry 
in late-time optical spectra of SNe~Ia. 

This paper is organized as follows.
In \S 2, the observational data of SN~2003hv and other SNe~Ia are summarized. 
In \S 3,  our model and the method of multi-dimensional spectrum synthesis are 
presented. 
The results are shown in \S 4 \& 5; 
\S 4 focuses on the geometry of SN 2003hv, while 
the implications for SNe Ia in general are presented in \S 5. 
The paper is closed with a discussion and our conclusions in \S 6. 
Future observing strategies to further constrain the explosion mechanism are 
also presented in \S 6.

\section{The Sample of SN Ia Late-Time Spectra}
The data set compiled for this work are summarized in Table~1. 
For SN~2003hv we focus on the optical spectrum at $\sim 320$ days after maximum brightness,  
taken with the {\it VLT} telescope equipped with FORS1, 
as reported by Leloudas et al. (2009). There are two more late-time spectra 
at $\sim 110$ and $143$ days after maximum light (Leloudas et al. 2009), 
which are briefly discussed in \S 6. 
A NIR spectrum of SN~2003hv 
was obtained on 2004 October 6 with the {\it Subaru} telescope 
equipped with CISCO/OHS (Motohara et al. 2006), 
394 days after maximum brightness. 
Leloudas et al. (2009) generated a combined optical-NIR 
spectrum mapped to 358 days 
by scaling the flux in the spectra mentioned above 
by interpolating optical and NIR photometric points. 
We also use a Mid-IR spectrum taken by Gerardy et al. (2007), 
with the {\it Spitzer} space telescope at 358 days. 

Optical late-time spectra of other SNe~Ia are compiled from the 
{\it SUSPECT} 
database\footnote{The Online Supernova Spectrum Archive, "SUSPECT", 
is found at http://bruford.nhn.ou.edu/$\sim$suspect/index1.html .}. 
We selected data which satisfy two criteria: 
(1) the spectrum is taken at least 100 days 
after maximum brightness, and 
(2) the spectrum covers the wavelength range 
$\sim 7,000 - 7,500$~\AA\ (for this criterion, see \S 4 \& 5). 
References for the data compiled from the {\it SUSPECT} 
database are listed in Table~1. 
Adding to this, 
we have also included SN 2000cx at the later epoch (+ 360 days) from 
Sollerman et al. (2004), and SN 2001el from Mattila et al. (2005).

\section{Multi-Dimensional Spectrum Synthesis}

\subsection{Model}

\begin{deluxetable*}{llllllll}
 \tabletypesize{\scriptsize}
 \tablecaption{Model Parameters\tablenotemark{a}
 \label{tab:tab2}}
 \tablewidth{0pt}
 \tablehead{
   \colhead{$v_{\rm off, ECAP}$}
 & \colhead{$v_{\rm off, LD}$}
 & \colhead{$v_{\rm ECAP}$}
 & \colhead{$v_{\rm HD}$} 
 & \colhead{$v_{\rm LD}$}
 & \colhead{$M_{\rm ECAP}$}
 & \colhead{$M_{\rm HD}$}
 & \colhead{$M_{\rm LD}$}
}
\startdata
3,500 & $-$1,500 & 3,000 & 5,000 & 10,000 & 
0.1 & 0.1 & 0.2
\enddata
\tablenotetext{a}{The units for velocities and masses are 
km~s$^{-1}$ and $M_{\odot}$, respectively. 
}
\end{deluxetable*}

\begin{figure}
\epsscale{1.0}
\plotone{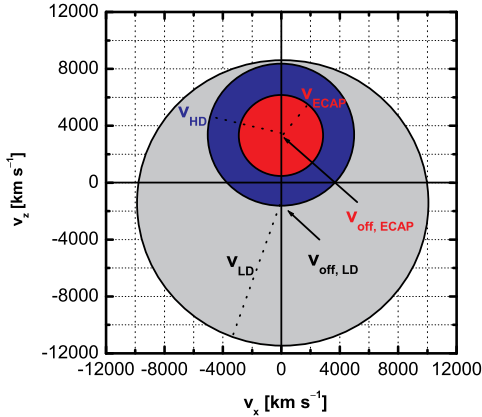}
\caption
{The kinematic model explored in the present study. 
The inner ejecta consist of three characteristic zones; the
ECAP region (red), High-Density (HD) $^{56}$Ni-rich region 
(blue), and Low-Density (LD) $^{56}$Ni-rich region (gray). 
The model is shown for $v_{\rm off, ECAP} = 3,500$ km s$^{-1}$ 
(the off-set of the ECAP and the HD regions) and 
$v_{\rm off} = -1,500$ km s$^{-1}$ (Tab. 2).   
\label{fig:fig1}}
\end{figure}

Rather than working with multi-dimensional, detailed hydrodynamic 
and nucleosynthesis models, 
we calculate nebular spectra for a simplified kinematic model, and try to 
obtain constraints on the model parameters from the 
late-time observations of SN 2003hv. 
In the model, we have three zones (Fig. 1); 
the off-set region filled with neutron-rich, Fe-peak elements produced 
with electron capture reactions (the ECAP zone), 
the off-set, relatively high density region filled with $^{56}$Ni 
(the HD zone), and the extended, relatively low density region filled 
with $^{56}$Ni (the LD zone). 
The model parameters are the following: 
\begin{itemize}
\item {\bf $v_{\rm off, ECAP}$:} 
The off-set velocity of the ECAP- and the HD-zones, 
with respect to the zero-velocity center of the ejecta. 
It is assumed that the off-set velocity is the same for 
the ECAP- and the HD-zones for simplicity. 
\item {\bf $v_{\rm off, LD}$: } 
The off-set velocity of the LD zone, with respect to the 
zero-velocity center. 
\item {\bf $v_{\rm ECAP}$, $v_{\rm HD}$, $v_{\rm LD}$: } 
The outer velocities of the ECAP-, HD-, and 
the LD-zones, respectively. 
The distribution of the elements and the density is assumed to be 
homogeneous within each region, 
with spherical symmetry with respect to the off-set position 
(but the HD/LD  zone has a hole corresponding to the ECAP/HD zone).
\item {\bf $M_{\rm ECAP}$, $M_{\rm HD}$, $M_{\rm LD}$: } The mass in each region. 
\end{itemize}

In each zone, the initial abundance is set as
follows; In the ECAP-zone, 10\% of the mass is in $^{56}$Ni,  and the
remaining 90\% is in stable $^{58}$Ni. The other regions
are assumed to  fully consist of $^{56}$Ni. 
$^{56}$Ni decays to $^{56}$Co and then to $^{56}$Fe, and thus the synthetic 
spectra are a mixture of forbidden lines of Ni, Co, and Fe. 
The masses are varied so that the model can roughly reproduce the
total flux as well as the fluxes of the emission lines of interest, 
as compared with the late-time spectra of SN 2003hv. The outer velocities are
set so that the predicted line widths are consistent with the observed
widths. The off-set velocities affect the wavelength centers of
various emission lines.  

Note that this model represents only the inner regions of 
the ejecta, and does not include the outer regions (i.e., 
the regions dominated by intermediate mass elements), 
since the outer regions contribute little 
to the late-time emission. 
Detailed model fit to all the spectral features is beyond the scope 
of this paper, as is evident from the simplifications in our model.
Also, the ejecta probably have a 
clumpy structure as well in realistic multi-D explosions, while 
such an effect is not included in the present study 
(see, e.g., Leloudas et al. 2009). 

The model is constructed to represent the  main features of
hydrodynamic explosion models,  especially of the delayed-detonation
model. The initial deflagration produces neutron-rich Fe-peak
elements with electron capture reactions,  and then $^{56}$Ni as the
density decreases at the deflagration front 
(e.g., Nomoto et al. 1984; Iwamoto et al. 1999; Brachwitz et al. 2000).  
We therefore regard the ECAP-zone and
the HD-zone as the products of the early deflagration phase 
(but see \S 6 for another scenario to create the off-center ECAP-zone). 
The subsequent
detonation produces a relatively low-density  $^{56}$Ni-rich region 
(the LD-zone). The distribution of the detonation products is expected
to be more or less spherical  (R\"opke \& Niemeyer 2007b; 
Kasen et al. 2009; Maeda et al. 2009c). We still allow for a small off-set  (opposite to
the deflagration) for the LD-zone, since  the detonation may
well be stronger in the direction where a larger amount of fuel  is
left after the deflagration phase.

\subsection{Multi-Dimensional Spectrum Synthesis}
The input model is mapped onto a three-dimensional Cartesian grid, with $51^{3}$ zones. 
Three-dimensional nebular spectrum synthesis calculations are then performed, 
using the $\gamma$-ray 
transfer module (Maeda 2006a) and the nebular spectrum synthesis 
module (Maeda et al. 2006c) 
of the {\it SAMURAI} code -- the {\it 
SupernovA MUlti-dimensional RadIation transfer} code\footnote{The {\it SAMURAI} 
is a compilation of 3D codes adopting Monte-Carlo methods to compute the high-energy 
light curve and spectra (Maeda 2006a), optical bolometric light curve (Maeda et al. 2006b), 
and optical spectra from early (Tanaka et al. 2006, 2007) to late phases (Maeda et al. 2006c).}. 
As a reference, we have also performed one-dimensional nebular 
spectrum synthesis for the classical, deflagration model W7 (Nomoto et al. 1984). 

Details of the calculation method are given in Maeda (2006a) 
(for $\gamma$-ray transport), Mazzali et al. (2001) 
(for one-zone nebular spectrum synthesis), and 
Maeda et al. (2006c) (for multi-dimensional spectrum synthesis). 
Good reviews on the nebular spectrum synthesis are given by 
e.g., Axelrod (1980), Ruiz-Lapuente \& Lucy (1992), 
Kozma \& Fransson (1998ab), and Liu et al. (1998). 

The $\gamma$-rays emitted by the decay chain 
$^{56}$Ni $\to$ $^{56}$Co $\to$ $^{56}$Fe 
produce non-thermal high energy electrons mainly by Compton scatterings, 
and the electrons give rise to impact ionization and excitation. 
The energy is also lost in thermalization and heating of the ejecta, 
through interactions with thermal electrons. 
An additional energy input comes from positrons emitted by the radioactive 
decays, with a stopping length much shorter than $\gamma$-rays. 
In our calculations, it is assumed that the positrons are fully trapped on the 
spot. The deposition of $\gamma$-rays and non-thermal positrons 
is the dominant heating process in the SN ejecta. 

\begin{figure}	
	\epsscale{1.0}
	\plotone{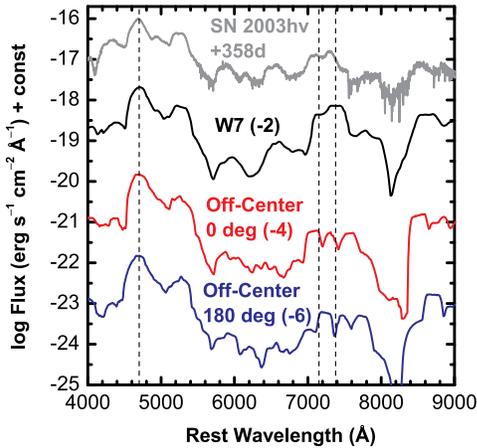}
\caption
{Synthetic late-time spectra of the SN~Ia models, as compared to 
the observed spectrum at day +320 (gray: 
dereddened with $E(B-V) = 0.016$~mag 
and  $R_{V} = 3.1$, the redshift corrected for the host galaxy 
recession velocity, and the flux 
calibrated to day +358. 
The model spectrum is calculated for 375 days after the explosion, and 
the flux is converted to the observed flux using 
the distance modulus $\mu = 31.37$. 
The W7 model of Nomoto et al. (1984) is a spherically symmetric model, 
and does not include any kinematic off-set. 
The model fluxes are shown in logarithmic scale, with offsets of 
-2 (W7: black), -4 (off-center, 0 deg: red), and -6 (off-center, 180 deg: blue).  
\label{fig:fig2}}

\end{figure}
\begin{figure}
	\epsscale{1.0}
	\plotone{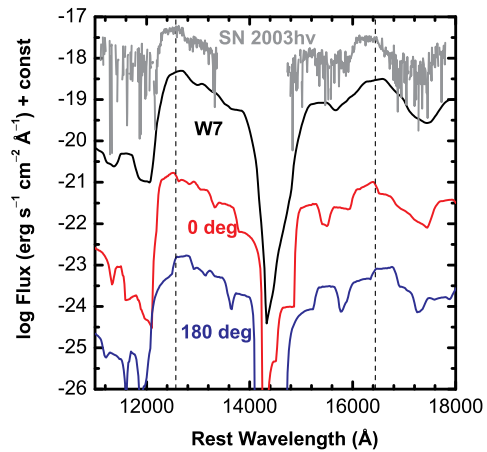}
\caption
{Same as Figure 2, but for NIR wavelengths. 
The original observed spectrum (Motohara et al. 2006) is again flux-scaled to 
day +358, 
and compared to the synthetic spectra at 375 days 
after the explosion.  The model fluxes are shifted by the same amounts as in Figure 2. 
\label{fig:fig3}}
\end{figure}

\begin{figure}
	\epsscale{1.0}
	\plotone{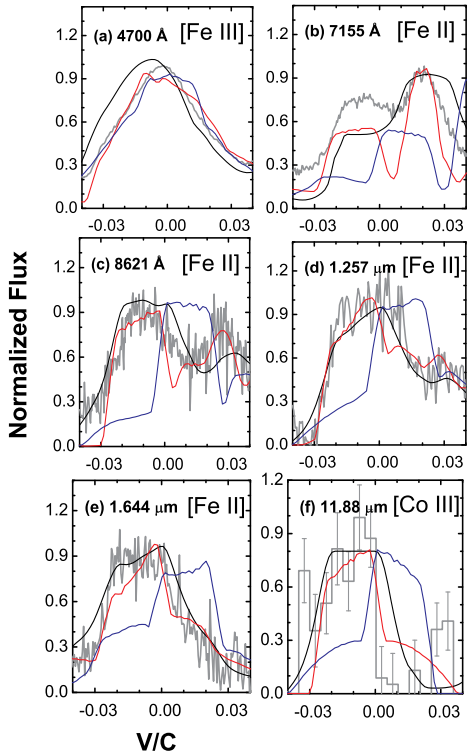}
\caption
{
Comparison between synthetic and observed line profiles.
The observations are shown in gray (see Table 1 for references). 
The black line is the W7 model artificially shifted to 
the blue by 2,800 km~s$^{-1}$. 
Our off-set model is shown for the observers in the direction of the off-set of 
the ECAP-zone (red) and in the opposite direction (blue). 
Unlike the models in Figures~2 and 3, the observed profiles are compared to 
the synthetic spectra at 333 days after the explosion in the optical (day +320) 
and at 410 days after the explosion in NIR (day + 398), in order to 
avoid possible evolutionary effects in the line profiles. The flux 
is arbitrarily normalized in this comparison.  
\label{fig:fig4}}
\end{figure}

At late phases, 
the heating is balanced by cooling from 
various thermal-collisionally excited 
emission lines, mostly forbidden. 
The non-thermal excitation is neglected in our calculations. 
The non-thermal impact ionization is balanced 
with recombination. As the density is low following the expansion, 
and the energy input is not dominated by thermal photons, the resulting ionization stages and 
level populations are generally not in Local Thermodynamic Equilibrium (LTE). 
For a given (thermal) electron temperature ($T_{\rm e}$), 
the ionization-recombination balance is solved (without photoionization 
in our calculations; see Sollerman et al. 2004), providing the 
electron density ($N_{\rm e}$) and ionization stages for each element. 
Once the ionization stage and the electron density are given, 
detailed balances for level populations are solved for each ion, 
under the condition of global heating-cooling balance. 
The balance then provides the electron temperature, 
as well as the resulting emission-line spectrum. 
In our numerical calculations, the ionization balance and the detailed balance 
for level populations are solved iteratively, until the electron density and 
temperature converge at every mesh point. 

The spectrum is sampled into 19 zenith angles, divided by 10 degrees each. 
The wavelength resolution in sampling the final spectrum corresponds to 
$\sim 450$~km~s$^{-1}$, roughly the same width as the spatial resolution 
of the input model. Note that any structure in the synthetic spectrum 
below this resolution is not reliable.

\section{SN 2003hv}

\subsection{Model Overview}

Figures 2 and 3 show synthetic spectra for the W7 and our off-set
model (Fig.~1), in the optical (Fig.~2) and in the NIR (Fig.~3), at 375 days
after the explosion.  In these figures, we show the observed spectrum
at 320 days (optical) and  398 days (NIR) after the maximum, but with
the flux calibrated to 358 days by Leloudas et al. (2009). 
Figure~4 shows the comparison of line profiles for some
selected emission lines. The model parameters are summarized in Table~2. 

Despite the simplification, the kinematic model can account for the
presence of the major emission lines in the late-phase spectrum of
SN~2003hv, and therefore also for those of SNe~Ia in general,  as do
the W7 model (see Axelrod 1980, who showed that the  main features of a
SN~Ia nebular  spectrum can be modeled by a simple, homogeneous
$^{56}$Ni-rich sphere).  
The observed flux, especially in the NIR, is
smaller than that predicted by the W7 model,  and consistent with
that in the kinematic model. This means that the mass of $^{56}$Ni
is $\sim 0.3 M_{\odot}$ in SN~2003hv, smaller than the typical value
in SNe~Ia  ($\sim 0.5 - 0.6 M_{\odot}$, e.g., Mazzali et al. 2007) and
than the W7 value ($\sim 0.6 M_{\odot}$).  This value is roughly
consistent with that estimated from the peak luminosity  ($\sim 0.4
M_{\odot}$) (Leloudas et al. 2009). 
We note that this is also related to relative contributions from 
different zones in our model, indicating that the W7 model has 
too much high density, low temperature materials to be consistent 
with the NIR flux.

\begin{deluxetable}{llll}
 \tabletypesize{\scriptsize}
 \tablecaption{Typical Values in Model Output\tablenotemark{a}
 \label{tab:tab3}}
 \tablewidth{0pt}
 \tablehead{
 \colhead{Zone}
 &  \colhead{$T_{\rm e}$}
 & \colhead{$\log N_{\rm e}$}
 & \colhead{Fe$^{+}$ Fraction\tablenotemark{b}}
}
\startdata
ECAP       &  2,000 & 5.7 & 90\%\\
HD         &  7,000 & 5.4 & 30\%\\
LD         & 11,000 & 4.8 & 5\%
\enddata
\tablenotetext{a}{Obtained as a result of spectrum synthesis 
for 375 days after the explosion. 
The units for $T_{\rm e}$ (the electron temperature) 
and $N_{\rm e}$ (the electron density) are in cgs. }
\tablenotetext{b}{The fraction of Fe$^{+}$. 
The remaining fraction is in Fe$^{++}$.}
\end{deluxetable}

In Figures 2 and 3, the expected rest wavelengths of some selected
emission lines are marked by lines:  [Fe~III] blend at 4700\AA, 
[Fe~II]~$\lambda$7155, and [Ni~II]~$\lambda7378$ in the optical (Fig.~2),  
[Fe~II]~1.257$\micron$ and [Fe~II]~1.644$\micron$ in the NIR (Fig.~3).  
As found by Motohara et~al.~(2006), the observed NIR emission features are
blueshifted as compared to those predicted  by a spherically symmetric
model like W7.  Leloudas et al. (2009) pointed out that 
[Fe~II]~$\lambda$7155 and [Fe~II]~$\lambda8621$ also show a similar
amount of  blueshift as compared to the expected wavelength.  
Furthermore, they suggested that [Fe~II]~$\lambda$8621 can be 
one of the best and cleanest lines to follow the geometry of the explosion. 

We here suggest that the feature at  $\sim 7300$~\AA\ is dominated by 
[Ni~II]~$\lambda$7378,
and then note that this feature is also blueshifted as compared to  the expected
position. At the same time, 
features in the blue part of the optical spectrum, e.g., at $\sim
4700$~\AA\ and $\sim 5250$~\AA,  do not show the blueshifts.  

This is more clearly seen in Figure 4, where the synthetic
spectrum for the W7 model  is artificially shifted in wavelength, by $2,800$ 
km~s$^{-1}$ to the blue, to fit the NIR features (see  Motohara et
al. 2006). The blueshift seen in [Fe~II]~$\lambda7155$ is also
explained by the same model. However, the features in the blue ($\sim
4700, 5250$~\AA) should also shift to the blue, and 
no longer fit the observed wavelengths. 

The inconsistency in the wavelengths is a strong argument against a
simple, bulk off-set model as the origin of the wavelength shift in
the NIR (and at $7155, 7378, 8621$~\AA).  Our 2D model (though computed in 3D) was
constructed to overcome this inconsistency.  As shown in Figures 2 - 4
(especially evident in Figure 4), the observed wavelengths of strong
lines,  in the optical, NIR, and Mid-IR wavelengths are all well explained by
our model, with the viewing orientation close to
the direction of the off-set. 

\subsection{Ionization/Thermal Structure and Line Shift}

\begin{deluxetable*}{llcrll}
 \tabletypesize{\scriptsize}
 \tablecaption{Optical and Infra-red Lines\tablenotemark{a}
 \label{tab:tab4}}
 \tablewidth{0pt}
 \tablehead{
 \colhead{Wavelength ($\micron$)}
 &  \colhead{Ion}
 & \colhead{Term}
 & \colhead{$E_{\rm u}$ (cm$^{-1}$)\tablenotemark{b}}
 & \colhead{Shift\tablenotemark{c}}
 & \colhead{Region\tablenotemark{d}}
}
\startdata
0.4658 & Fe III & $^{5}$D$_{4}$ - $^{3}_{2}$F$_{4}$& 21462.2 & No & LD\\
0.4701 & Fe III & $^{5}$D$_{3}$ - $^{3}_{2}$F$_{3}$& 21699.9 & No & LD \\
0.4734 & Fe III & $^{5}$D$_{2}$ - $^{3}_{2}$F$_{2}$& 21857.2 & No & LD \\
0.5262 & Fe II & a$^{4}$F$_{7/2}$ - a$^{4}$H$_{11/2}$& 21430.4 & No & LD \\
0.7155 & Fe II & a$^{4}$F$_{9/2}$ - a$^{2}$G$_{9/2}$& 15844.7 & Yes & HD \\
0.7378 & Ni II & $^{2}$D$_{5/2}$ - $^{2}$F$_{7/2}$ & 13550.4 & Yes & ECAP \\
0.8617 & Fe II & a$^{4}$F$_{9/2}$ - a$^{4}$P$_{5/2}$& 13474.4 & Yes & HD \\
1.257 & Fe II & a$^{6}$D$_{9/2}$ - a$^{4}$D$_{7/2}$& 7955.3 & Yes & HD \\
1.644 & Fe II & a$^{4}$F$_{9/2}$ - a$^{4}$D$_{7/2}$& 7955.3& Yes & HD \\
2.218 & Fe III & $^{3}$H$_{6}$ - $^{3}$G$_{5}$ & 24558.8 & No & LD \\
2.348 & Fe III & $^{3}$H$_{5}$ - $^{3}$G$_{5}$ & 24558.8 & No & LD \\
2.874 & Fe III & $^{3}_{2}$F$_{3}$-$^{3}$G$_{3}$ & 24940.9 & No & LD \\
2.904 & Fe III & $^{3}_{2}$F$_{3}$-$^{3}$G$_{3}$ & 25142.4 & No & LD \\
3.228 & Fe III & $^{3}_{2}$F$_{4}$-$^{3}$G$_{5}$ & 24558.8 & No & LD \\
4.114 & Fe II & a$^{6}$D$_{9/2}$ - a$^{4}$F$_{7/2}$& 2430.1 & Yes & HD + ECAP \\
4.606 & Fe II  & a$^{6}$D$_{5/2}$ - a$^{4}$F$_{5/2}$ & 2838.0 & Yes & HD + ECAP \\
4.888 & Fe II & a$^{6}$D$_{1/2}$ a$^{4}$F$_{3/2}$ & 2430.1 & Yes & HD + ECAP \\
5.339 & Fe II & a$^{6}$D$_{9/2}$ - a$^{4}$F$_{9/2}$ & 1872.6 & Yes & HD + ECAP \\
6.634 & Ni II & $^{2}$D$_{5/2}$ - $^{2}$D$_{3/2}$& 1506.9 & Yes & ECAP \\
7.350 & Ni III & $^{3}$F$_{4}$ - $^{3}$F$_{3}$& 1360.7 & Yes & ECAP \\
10.52 & Co II & a$^{3}$F$_{4}$-a$^{3}$F$_{3}$ & 950.51 & Yes & HD \\
10.67 & Ni II & $^{4}$F$_{9/2}$ - $^{4}$F$_{7/2}$& 9330.0 & Yes & ECAP \\
11.00 & Ni III & $^{3}$F$_{3}$ - $^{3}$F$_{2}$& 2269.6 & Yes & ECAP \\
11.88 & Co III & a$^{4}$F$_{9/2}$ - a$^{4}$F$_{7/2}$& 841.2 & Yes & HD + LD \\
12.72 & Ni II & $^{4}$F$_{7/2}$ - $^{4}$F$_{5/2}$& 10115.7 & Yes & ECAP \\
14.74 & Co II & a$^{5}$F$_{5}$-a$^{5}$F$_{4}$ & 4029.0 & Yes & HD \\
17.93 & Fe II & a$^{4}$F$_{9/2}$ - a$^{4}$F$_{7/2}$&2430.1 & Yes & HD + ECAP
\enddata
\tablenotetext{a}{This is not a complete line list; especially, in the optical 
here are only the lines discussed in the main text. The NIR and Mid-IR 
line list covers the strongest lines in the model. 
}
\tablenotetext{b}{The upper energy level of the transition.}
\tablenotetext{c}{Shift in the wavelength depending on the viewing angle.}
\tablenotetext{d}{Region which makes the predominant contribution.}
\end{deluxetable*}

These model results can be understood on the basis of the 
ionization and thermal structure.  Table~3 shows the typical electron
temperature, electron density, and ionization stage in each of the
zones. The ECAP-zone is in a low
ionization state (mostly singly-ionized species), because  of the small energy
input (inefficient ionization) and high density (efficient
recombination).  Despite the low ionization state, the electron density is
high (because of large material density).  The electron temperature is
accordingly low. In the LD-zone, the situation is the opposite.
It is in a high ionization state  (mostly doubly-ionized), 
the electron density is small and thus the electron
temperature is kept as high as $\sim 10,000$ K. 
The HD-zone has properties intermediate between these two zones. 

The characteristic strong emission lines in the synthetic spectrum 
are summarized in Table~4. 
The strong feature at $\sim 4,700$\AA\ is a blend of several 
[Fe~III] lines, all of which are 
formed under the high ionization and high temperature condition. 
The feature is thus dominated by emissions from the extended LD zone, 
which is distributed in an approximately spherical way. 
Therefore, the feature at  $\sim 4700$\AA\ 
does not show a large velocity shift. 
The [Fe~II]~$1.257~\micron$ and $1.644~\micron$, on the other hand, 
come from the low ionization and low temperature conditions, 
and thus are emitted mostly 
from the off-set HD-zone. Therefore, these NIR [Fe~II] lines 
show a large blueshift if viewed from the direction of the off-set. 
[Ni~II]~$\lambda$7378 is emitted from the ECAP zone (i.e., $^{58}$Ni), 
and show a large velocity shift. 
Contribution from radioactive $^{56}$Ni to this line is negligible at these 
late phases because of the short decay time scale.

The behavior of other [Fe~II] lines in the optical range can be understood
in the same manner, but is complicated by a competition between the
ionization and temperature effects.  [Fe~II]~$\lambda\lambda7155$ and 8621 
show the velocity shift, as these are mostly emitted from the HD-zone. 
This is due to the low ionization in this zone.  On
the other hand, the feature at $\sim 5250$~\AA\ does not show a large
shift (Fig.~2),  which is not interpretable by the ionization effect alone. 
The strongest line in this feature is 
[Fe~II]~$\lambda$5262. The excitation temperature for this line is high,
and thus the high temperature in the LD-zone is preferred,
despite the small amount of Fe$^{+}$ present there. 
Also, the contribution of
[Fe~III]~$\lambda$5270 from the LD-zone is not
negligible. As a result, the large contribution to the emission
feature at $\sim 5250$~\AA\ comes from the LD-zone. This
explains why this feature does not show a large velocity shift. 

[Co~III]~$11.88~\micron$ is a ground state
transition with an excitation temperature of only $\sim 1,000$~K, much
lower than the temperatures in either the HD or LD-zone. 
In addition, the fraction of the "doubly"-ionized Co is similar between
these two zones (unlike the singly-ionized Co). 
The LD and HD zones thus provide
comparable contributions to the [Co~III]~$11.88~\micron$ feature. 
As a result, the line profile is a
combination of a broad component  centered at the rest wavelength
and a relatively narrow component whose central wavelength is
blueshifted if viewed from the off-set direction. Because of the
strong narrow component,  the line as a whole shows a velocity
shift, depending on the viewing direction.

\section{Signatures of Explosion Asymmetry in SNe Ia in General}

\begin{figure}
	\epsscale{0.8}
	\plotone{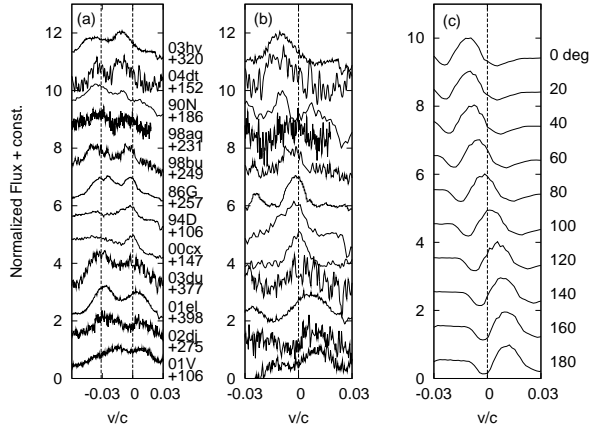}
\caption
{Analysis of the [Ni~II]~$\lambda$7378 line profiles in the 12 SNe~Ia. 
The velocity 
is set assuming that the rest wavelength is at 7378~\AA. 
(a) Observed line profiles. The rest wavelengths of [Fe~II]~$\lambda$7155 
and [Ni~II]~$\lambda$7378 are shown by dotted lines. 
(b) [Ni~II]~$\lambda$7378 in observations, after removing the 
underlying continuum (or possible other lines) as much as possible (see main text). 
(c) Synthetic line profiles of the [Ni~II], 
depending on the viewing orientation. 
\label{fig:fig5}}
\end{figure}

In \S 4, we have shown that the somewhat 
puzzling features in the late-time spectra of SN~2003hv,  i.e., some 
lines showing a blueshift while others showing no velocity shift, 
can be explained by a geometrical effect. 
In this section, we expand the discussion in \S 4; we investigate the 
explosion geometry of SNe~Ia in general. 
We also present diagnostics of the 
geometry using NIR and Mid-IR lines (Tab. 4), for existing as well as for 
future observatories such as {\it JWST} 
(James Webb Space Telescope) or {\it SPICA} 
(SPace Infrared telescope for Cosmology and Astrophysics mission).

\subsection{[Ni~II]~$\lambda$7378 as diagnostics of the deflagration phase}

It has been believed that signatures of ejecta
asymmetry are not evident in the optical range (\S 1 and \S 5.2). 
Figure 5 shows the
feature around $\sim 7000 - 7500$~\AA,  i.e., [Fe~II]~$\lambda7155$ (with
some contribution from [Fe~II]~$\lambda7171$)  and [Ni~II]~$\lambda$7378,
for the 12 SNe~Ia (Table~1; see below). 

Contrary to the earlier expectation,  we find that there is indeed a
probable signature of ejecta asymmetry.  
As shown in Figure 5, the
feature shows a variation in the central wavelength  
(after removing the host galaxy redshift). 
In many cases, the shift is larger than
$1,000$~km~s$^{-1}$, which is too much to be due to 
a peculiar motion of the SN progenitor with respect to the host center. 
The [Fe~II]~$\lambda7155$ and [Ni~II]~$\lambda7378$ 
always show  a similar degree of the wavelength shift. 

Both blueshifts and redshifts exist, and 
the shift does not appear to correlate with the age of the SN 
(see \S 6 for further discussion). 
These observed characteristics suggest that the shift is not caused either by 
radiation transfer effects or 
by other unidentified emission lines. Thus, the mounting evidence is that the 
variation of the central wavelengths represents 
a real variation in the line-of-sight velocity of the region 
emitting [Fe~II]~$\lambda$7155 and [Ni~II]~$\lambda$7378. 

As shown in \S 4, these lines are emitted from deep parts of the ejecta: 
[Fe~II]~$\lambda$7155 from the HD-zone and [Ni~II]~$\lambda$7378 
from the ECAP-zone in our model. 
We point out that especially the [Ni~II] 
should preserve important information on 
the explosion dynamics, since the ECAP-zone is attributed to the region 
created by the initial deflagration (or, the detonation passing through the central region 
of the white dwarf if the initial deflagration is very weak; \S 6), 
with efficient electron capture reactions.  
This is supported by the relatively narrow width ($\lsim 3,000$~km~s$^{-1}$) of 
the [Ni~II] line. 

One may argue that the central wavelength of the
[Ni~II]~$\lambda$7378 could apparently shift to the blue  because the
blue wing of this feature is contaminated by the  [Fe~II] (and by
[Ca~II]~$\lambda\lambda$7291, 7307,  although it does not likely
contribute much\footnote{We have not included Ca in our model. 
The W7 model predicts that the contribution from 
[Ca~II]~$\lambda$7291, 7307 is about 10\% (Leloudas et al. 2009). Also, the W7 model 
predicts broad [Ca~II] which is not compatible to the observed relatively 
narrow emission features.}).  
To check this, we  tried to obtain intrinsic
profiles of the [Ni~II] line by subtracting the possible contamination 
of other lines, especially in the blue wing (Fig. 5). Practically, we
assumed a (pseudo)-continuum connecting the flux minima on both side 
of $7,380$~\AA\ (the blue minimum corresponds to the  valley between
the [Fe~II] and [Ni~II]). 
This continuum flux generally decreases
towards the red, and thus subtracting it could
result in a shift of the peak wavelength to the red to some
extent. However, we found that this effect is small, as compared
to the observed variation. 

Another possible concern is the simplified abundance
distribution in our model, in which the HD and LD-zones are
assumed to contain no stable Ni. Even without electron captures
realized in the early deflagration phase,  some amount of $^{58}$Ni
should be produced in the region where $^{56}$Ni is the dominant
burning product. The typical mass fraction of $^{58}$Ni in such a region
is $\sim5\%$ for solar metallicity  (e.g., Iwamoto et al. 1999;
Timmes et al. 2003). To check the effect, we replaced 5\% of the material by 
$^{58}$Ni in the model and repeated the calculations. We thereby
confirmed that the result is not defeated by this change, and that
[Ni~II]~$\lambda$7378 emitted from the HD/LD-zones is
negligible  as compared to that from the ECAP-zone. 

\begin{figure}
	\epsscale{0.8}
	\plotone{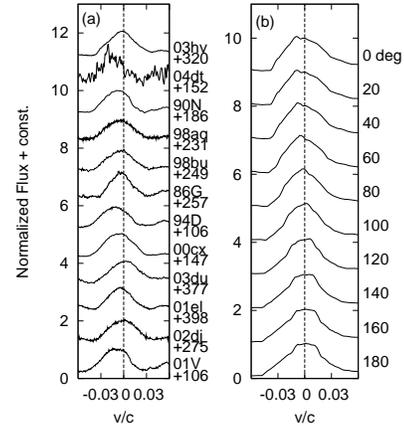}
\caption
{Analysis of the [Fe~III]~blend at 4700\AA\ 
in the 12 SNe Ia. The velocity 
is set assuming the rest wavelength of 4700~\AA. 
(a) Observed line profiles. The rest wavelengths of [Fe~III]~$\lambda$4701  
is shown by a line. 
(b) Synthetic line profiles of the [Fe~III] feature, 
depending on the viewing orientation. 
\label{fig:fig6}}
\end{figure}

The [Ni~II]~$\lambda7378$ line is not always strong in the  
observed spectra of SNe~Ia.  We find
that 12 SNe~Ia (Fig.~5) show this feature among the 20 SNe we investigated
(Table 1).  The number of SNe~Ia, showing the [Ni~II]~$\lambda$7378
stronger than the [Fe~II]~$\lambda$7155 is even smaller.  
Although  [Ni~II]~$\lambda$7378 
is mainly emitted from the ECAP-zone in our present calculations,   
the contribution from the HD/LD regions may not be negligible 
in case of larger $M_{\rm HD}$ and/or $M_{\rm LD}$. 
Also, with an increasing amount of $^{56}$Ni, [Fe~II]~$\lambda$7388 
and [Fe~II]~$\lambda$7452,  could hide the [Ni~II] feature. 
Thus, the observed diversity in the detection of the [Ni~II] 
may indicate that $M_{\rm ECAP}/(M_{\rm HD} + M_{\rm LD})$ 
varies among objects (e.g., Mazzali et al. 2007). 

\subsection{[Fe~III] blend as diagnostics of the outer region}

In Figure 6, we show line profiles of the same sample of SNe Ia, 
but centered at $4700$~\AA. 
Unlike the [Ni~II]~$\lambda$7378, there is no large variation. The absence of 
a significant wavelength shift has been a strong argument 
against any global asymmetry in 
SNe~Ia, since this feature is the strongest in SN~Ia nebular spectra 
and has thus naturally been used to infer the geometry. 

\begin{figure}
	\epsscale{0.8}
	\plotone{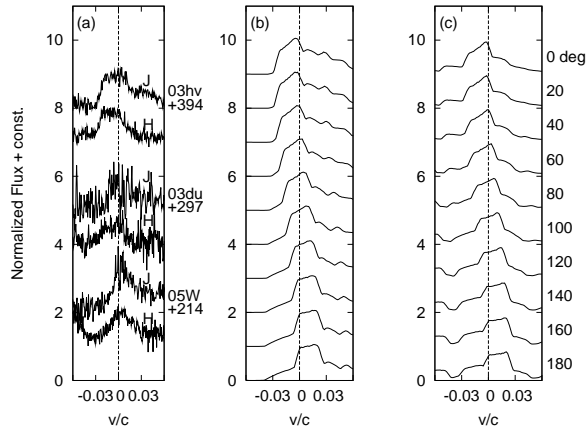}
\caption
{Analysis of NIR [Fe~II] line profiles in the 3 SNe Ia 
(H\"oflich et al. 2004; Motohara et al. 2006). The velocity 
is set assuming the rest wavelength of 1.257~$\micron$ (J-band) and 
1.6440~$\micron$ (H-band). 
(a) Observed line profiles. 
(b) Synthetic line profiles of [Fe~II]~1.257~$\micron$. 
(c) Synthetic line profiles of [Fe~II]~1.644~$\micron$. 
\label{fig:fig7}}
\end{figure}

This feature is a blend of several [Fe~III] lines, with the wavelength separations 
smaller than the typical line width (10,000 km s$^{-1}$); 
[Fe~III]~$\lambda$4658, [Fe~III]~$\lambda$4701, [Fe~III]~$\lambda$4734, 
[Fe~III]~$\lambda$4755, [Fe~III]~$\lambda$4769, and [Fe~III]~$\lambda$4778. 
As a result of the blending, the central wavelength of the feature is 
$\sim 4,700$\AA. Note that this central wavelength is not sensitive 
to underlying models as long as spherically symmetry is assumed, 
since the excitation temperature of these lines are all similar and thus 
the relative contribution is basically determined by the transition probabilities. 

With the central wavelength of $4,700$\AA, 
no redshifts are observed in these SNe~Ia. 
There are, on the other hand, some SNe~Ia showing small blueshifts in this feature. 
This argues against the idea that the wavelength shifts here are 
caused by a geometrical effect. We note that the SNe~Ia that show relatively 
large blueshifts ($\gsim 1,500$ km s$^{-1}$) 
are mostly young objects (the spectra taken before day +200) -- 
SNe 2004dt (+152), 1990N (+186), 1994D (+106), 2000cx (+147), and 
2001V (+106). 
The central wavelength of the 4,700\AA\ feature in 
spectra taken at $\gsim 200$ days is clustered at the expected 
wavelength ($\sim 4,700$~\AA). 
This is further discussed in \S 6. 
We conclude that the wavelength shift in the $4700$~\AA\ feature 
is not caused by geometric effects, but by either a 
radiation transfer effect or contamination from other lines, e.g., 
Mg~I]~$\lambda$4571.

The behavior that the [Fe~III] blend is always at $\sim$ 4,700\AA\ 
in sufficiently late phases ($\gsim$ +200 days) can be 
understood from our model results. Since this blend is emitted from the LD zone, 
it indicates that the LD zone is more spherically distributed than the inner 
ECAP and HD zones. 

\subsection{NIR [Fe~II] lines as a promising tool to investigate the geometry}

Figure 7 shows a similar analysis for the NIR (J and H) [Fe~II] features -- 
[Fe~II]~$1.257~\micron$ and $1.644~\micron$. 
As discussed in \S 4, these lines are mostly emitted from the 
HD-zone, and thus show a large variation as a function of 
the viewing angles. 

Not many observational data are available for these NIR lines, 
because of the difficulty raised by the OH airglow lines. 
In Figure 7 we show the data presented by Motohara et al. (2006). 
NIR spectra for two other SNe~Ia have been reported: 
SN 1991T (Bowers et al. 1997) and SN~1998bu 
(Spyromilio et al. 2004). The [Fe~II]~$1.644~\micron$ line in SN~1991T 
is almost symmetric 
with respect to the rest frame of the host galaxy, 
while SN~1998bu seems to show a small degree of blueshift. 

In principle, it is favorable to check the consistency of the wavelengths 
of various lines, as we have done for SN~2003hv. [Fe~II]~$1.644~\micron$ 
in SN 2003du seems to show some blueshift (H\"oflich et al. 2004; 
Motohara et al. 2006), although the Signal-to-Noise ratio (S/N) is not good. 
[Ni~II]~$\lambda$7378 in SN 2003du is only marginally detected (Fig. 5),  
and the poor S/N makes it difficult to determine the exact position 
of the center of the line. In SN~1998bu, [Ni~II]~$\lambda$7378 is relatively 
well identified with better S/N than in SN~2003du, and it shows $\sim 1,000$ 
km s$^{-1}$ of the blueshift, at least qualitatively in agreement with the 
NIR observation.  

\subsection{Mid-IR lines as a new probe}

\begin{figure}
	\epsscale{0.8}
	\plotone{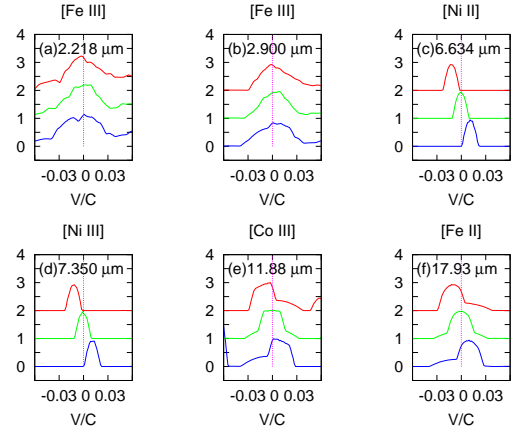}
\caption
{Analysis of Mid-IR line profiles. 
Synthetic line profiles are shown for various lines (a -- f; see Tab. 4). 
Three lines in each panel are for different viewing angles 
(0, 90, 180$^{\circ}$ from top to bottom; red, green, and blue, respectively). 
\label{fig:fig8}}
\end{figure}

\begin{figure*}
	\epsscale{1.0}
	\plotone{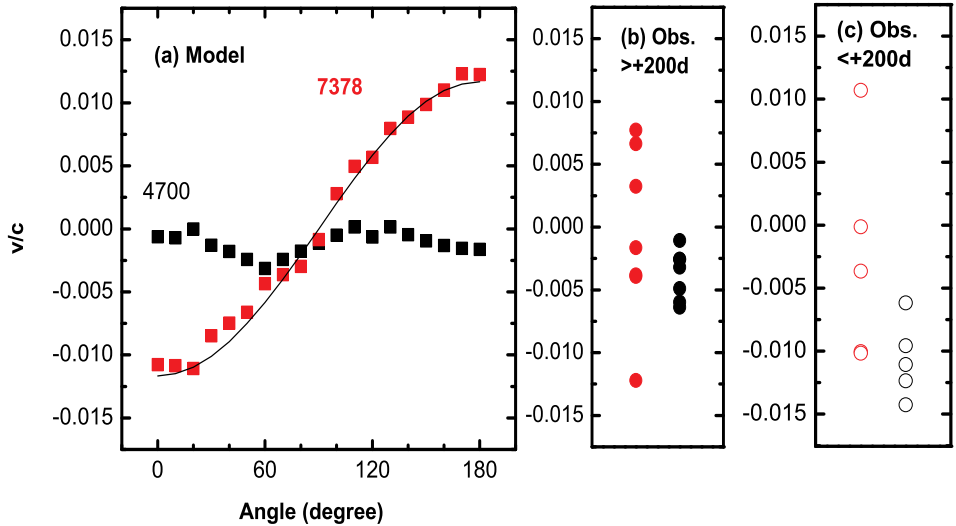}
\vspace{-1cm}
\caption
{(a) 
The central wavelength of [Ni~II]~$\lambda$7378 (red filled squares) and 
the [Fe~III] blend at 4700\AA\ (black filled squares) in the synthetic spectra 
as a function of the viewing angle. Note that the binning in the synthetic spectrum 
is about $0.0015 c$, and fluctuations at this level 
can simply be numerical noise. 
The angle dependence of [Ni~II]~$\lambda$7378 
is consistent with $v/c \sim 0.012 \cos(\theta)$ as 
expected. On the other hand, the 4700\AA\ feature 
is not sensitive to the viewing angle.
Panels (b) and (c) show the line shift 
derived for the 12 SNe Ia. 
The SNe are divided according to their phase:  
(b) SNe after day +200, and (c) SNe before day +200. 
The red symbols are for [Ni~II]~$\lambda$7378, 
while the black ones are for the [Fe~III] blend at 4700\AA. 
For both model and observed spectra, 
the central wavelength is defined as the wavelength 
with 50 \% of the total line luminosity on either side. 
\label{fig:fig9}}
\end{figure*}

In the mid-IR wavelength range, there are various isolated forbidden
lines of Fe, Co, and Ni.  There are also lines from intermediate
mass elements such as [Ar~II] and [Ar~III]  (Gerardy et al. 2007),
which are not included in our model.  
For these longer wavelengths,
lines tend to be more isolated, and thus hardly affected by a line
blending. The lines in the mid-IR therefore provide a direct view on the
geometry of the emitting region. 

In Figure~8, some selected lines synthesized in our model  are
shown. Emission features at $\sim 2.2~\micron$  and $\sim 2.9~\micron$
are both dominated by [Fe~III] with an excitation temperature of 
$\sim 30,000$~K. Thus, these lines behave in the same way as 
the [Fe~III] blend at 4,700\AA; virtually no wavelength shift is seen,
irrespective of the viewing orientation. 

The behavior of [Ni~II]~$6.634~\micron$ is basically the same as in 
[Ni~II]~$\lambda$7378.  This line provides a very strong diagnostics on
the geometry of the initial deflagration phase, even better than 
[Ni~II]~$\lambda$7378: 
(1) the excitation temperature is only  a few
1,000~K (thus no doubt it is emitted from the ECAP-zone where not much
heating is taking place), and 
(2) there is no blending with other lines. 
These effects are already taken into account in the analysis of
[Ni~II]~$\lambda$7378, but [Ni~II]~$6.634~\micron$ can provide the
the same diagnostics in a more model-independent way. 

On the other hand, the analysis of [Ni~III]~$7.350~\micron$ may 
be complicated by the
competition between the high ionization and low excitation
temperature.  In our model, it is dominated by the emission from
the ECAP-zone.  We have confirmed that adding a representative
amount of $^{58}$Ni in the  HD/LD-zones does not affect our
result (\S 5).  

[Co~III]~$11.88~\micron$ has already been discussed in \S 4, in the
context of SN~2003hv.  From Figure~8, it is seen that this line has
two components; a broad symmetric feature  arising from the LD zone 
and a narrow component from the HD-zone.  
The former is always at the rest wavelength, but the latter shows a
variation  in the central wavelength as a function of the viewing
angle.  
The relative strength may well vary depending on
$M_{HD}$/$M_{LD}$.  In the present model, the ratio is set
by the requirement that the fluxes in the blue part of  the optical range
(mainly emitted from the LD-zone) and in the NIR  (mostly from the HD zone)
should be consistent with the observed spectra of SN 2003hv. 

[Fe~II]~$17.93~\micron$ behaves in a way similar as [Fe~II]~$1.644~\micron$.  
A large variation in the central wavelength is seen. One
important difference is that we do see a  contribution from the ECAP-zone.
Because of the low  excitation temperature of 
[Fe~II]~$17.93~\micron$, 
this comes from a small amount of the decay product of $^{56}$Ni
in the ECAP-zone where the electron temperature is low. 
As a result, this line does not  show a
flat-topped profile, as seen in [Fe~II]~$1.644~\micron$  
(\S 6 for further discussion).

\section{Discussion and Conclusions}

\subsection{Summary}

In this paper, we have investigated the geometry of the innermost
region of  SNe Ia, which should provide information
on the explosion mechanism.
We have shown that some enigmatic properties observed in
the late-time optical, NIR, and Mid-IR spectra of SN Ia
2003hv -- that some emission lines show blueshift while others do
not -- can in fact be naturally explained by a simple kinematic model, 
if the viewing angle is close to the direction of the off-set. 
In this model, the high density regions 
(the ECAP-zone dominated by stable Fe-peak
elements, and the HD-zone dominated by $^{56}$Ni) are
off-set with respect to the center of the expansion by $\sim 3,500$
km s$^{-1}$, and are surrounded by a less asymmetric,  low density
$^{56}$Ni-rich region (the LD-zone). 

Table 4 summarizes the expected qualitative behavior of strong emission lines in
late-time nebular SN Ia spectra.  Forbidden lines from ions with high
ionization stage and/or high excitation temperature are mostly
emitted from the outer, "spherical" LD-zone.  Such lines
show little variation in the central wavelength, irrespective of the
viewing angle.  On the other hand, lines from ions with low
ionization stage and/or  low excitation temperature are generally
emitted from the inner, "off-set" ECAP- or HD-zones. 
These lines show a strong variation in the central 
wavelength as a function of the viewing angle.  

\begin{figure}
	\epsscale{1.0}
	\plotone{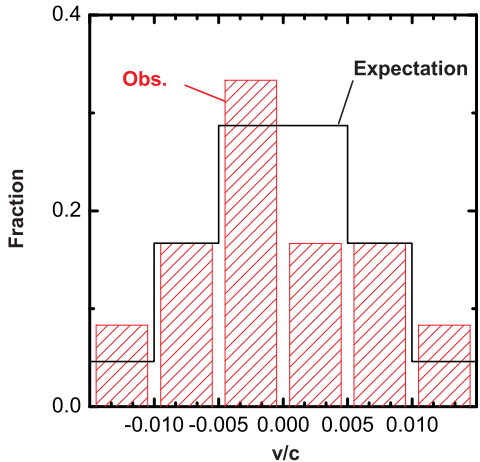}
\caption
{The number distribution of SNe Ia as a function of 
the central wavelength of the [Ni~II]~$\lambda$7378 
feature. The observed number distribution, using 12 SNe~Ia, is shown by 
shaded bars. 
The expected distribution from the present model (as a function of the 
viewing angle) is shown by the line. 
\label{fig:fig10}}
\end{figure}

Encouraged by this finding, we have investigated available optical 
late-time spectra of SNe Ia compiled mostly from the {\it SUSPECT} archive. 
Figure 9 summarizes our results from the model calculations, 
and the measurements in the 12 observed SNe~Ia sample. 
From the model, we see that the variation in the central wavelength of 
[Ni~II]~$\lambda$7378 basically follows the simple kinematic expectation. 
Note, however, for a precise prediction of the behaviors of emission lines, 
relative contributions from the different zones have to be considered, 
which further depend on the ionization structure, 
the electron temperature and the density. 
For [Ni~II]~$\lambda$7378, the effects of these details turn out to be 
relatively small, but some lines are indeed affected by these complications (\S 5). 
We have provided an explanation on the observational behavior that 
[Ni~II]~$\lambda$7378 shows the diversity in the central wavelength while 
[Fe~III] blend at 4,700\AA\ does not: they trace different zones, and 
the inner ECAP and HD zones have the large off-set while the outer LD 
zone does not. 

We suggest that [Ni~II]~$\lambda$7378 to be one of the few lines 
at optical wavelengths to provide 
strong diagnostics of the innermost region of SNe Ia. 
Among the 20 SNe Ia, about half of them (12 SNe) show a probable 
detection of this feature. 
Similar diagnostics can be obtained by [Fe~II]~$\lambda7155$ as well. 
Thus, the spectra in the wavelength range of $\sim 7,000 - 7,500$\AA\ 
should provide 
useful diagnostics of the explosion asymmetry. 
With present instrumentation, these features can be more easily observed 
than other strong indicators in the NIR and Mid-IR regime. 
A late-time spectrum of an overluminous SN Ia 
2006gz (Hicken et al. 2007) also shows a hint of the possible blueshift in these  
features (Maeda et al. 2009b). 
However, the spectrum is very noisy and the identification is not secure. 
Also, the spectrum was peculiar in a sense that it did not show the strong features 
in the blue (i.e., at $\sim 4,700$\AA). For these reasons, we have not included 
SN 2006gz in our analysis, despite the possible importance of asphericity in this object 
(Hillebrandt et al. 2007; Sim et al. 2007; Maeda \& Iwamoto 2009a).

The geometry we derived could naturally arise in SN Ia explosions. 
The initial deflagration may proceed in a very asymmetric way, 
while the subsequent detonation is expected to leave $^{56}$Ni 
more spherically distributed. 
An important question is whether the geometry suggested for 
SN 2003hv represents a special case or 
a general structure of SN Ia explosions. Figure~10 shows 
the distribution of 12 SNe Ia, as a function of the velocity 
shift in [Ni~II]~$\lambda$7378. Although the sample 
is small, 
the general agreement 
between the observations and the expectation assuming 
the geometry of SN 2003hv as a generic feature of SN Ia explosions, is very 
encouraging.  From this analysis, it seems that
SN 2003hv may not be very different from 
other SNe~Ia, and its largest velocity shifts can be attributed to the 
viewing angle effect. 

\subsection{Temporal Evolution}

\begin{figure}
	\epsscale{1.0}
	\plotone{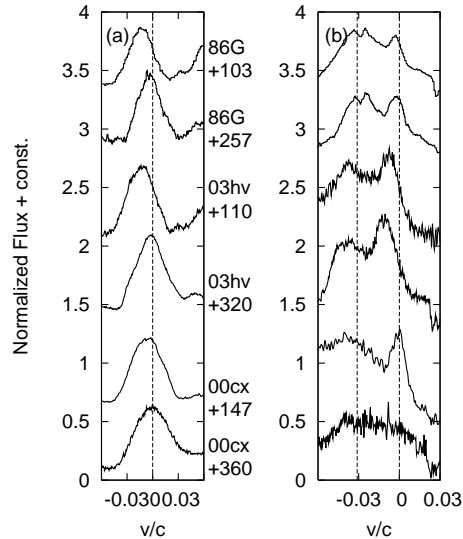}
\caption
{Examples of the temporal evolution in 
(a) the region around the [Fe~III] blend at 4700\AA\ and 
(b) that around [Ni~II]~$\lambda$7378. 
The data are taken from Cristiani et al. (1992) 
for SN~1986G, Leloudas et al. (2009) for  
SN~2003hv, and Li et al. (2001) and Sollerman et al. (2004) for SN~2000cx. 
\label{fig:fig11}}
\end{figure}

We have found that the line center of [Ni~II]~$\lambda$7378 does not 
correlate with the epoch at which the spectrum was taken. 
Therefore, the line velocity shift should reflect the intrinsic geometry and 
variation in the viewing angle. On the contrary, the central wavelength of the 
[Fe~III] blend at 4700\AA\  does show a clear correlation with the epoch; 
The line always shows a blueshift before $\sim 200$ days, 
while there is no significant wavelength shift in spectra taken after $\sim 200$ days 
(Fig. 9). 
This is  exemplified by the temporal evolution of the nebular spectra, available for 
a few SNe Ia (Fig. 11): 
Thus, we suggest that the [Fe~III] traces the intrinsic geometry 
only after $\sim 200$ days, and that the outer, relatively low density 
region is less aspherical than the innermost region. 

Figure~11 shows that there is also diversity in the temporal evolution of 
[Ni~II]~$\lambda$7378. 
In SNe~1986G and 2003hv, [Ni~II]~$\lambda$7378 persisted over a long
period, while it only shows a transient feature in SN~2000cx. This may indeed be consistent with 
the model calculation; in our calculation, the strength of [Ni~II]$\lambda$7378 
is largely reduced at $\sim 350$ days (Figs. 2 \& 4), following the temperature drop and 
the so-called infrared catastrophe in the ECAP zone (see also \S 6.3). 

The central wavelength of the [Ni~II] line does not evolve in SN~1986G, while
it shows a somewhat larger blueshift in the later phase in  SN~2003hv. 
The difference is likely related to
the distribution of different species, and the resulting  $\gamma$-ray
deposition and potentially the positron escape. 
An increasing degree of the blueshift in SN~2003hv as a function of time might 
indicate that $^{56}$Ni is distributed inhomogeneously within the
ECAP zone, being relatively concentrated at the high velocity 
"edge" of this zone (e.g., $\sim 2,000 - 3,500$ km s$^{-1}$).  In such a scenario, 
$\gamma$-rays irradiate the whole ECAP zone early on, while later the
positrons heat only the relatively "high-velocity" region.

The number of SNe~Ia with a good temporal coverage at late phases
is still small, 
and it is strongly encouraged to perform multi-epoch spectroscopy. 
Such intensive observations are useful to 
understand (1) the detailed distribution of $^{56}$Ni, and 
(2) detailed thermal processes (e.g., $\gamma$-ray deposition, 
positron escape, IR catastrophe) in the SN Ia nebulae. 
It is also important to investigate (3) a possible evolution 
in the line profiles if we are to use them to investigate the details 
of the explosion physics (see \S 6.3 for more details). 

\subsection{Future Perspectives}

The conclusion in the present paper that the late-time nebular spectra 
can be used to investigate the geometry of ejecta, thus the explosion 
mechanism of SNe Ia, provides a new strategy to clarify the nature of 
SN Ia explosions. Presently, the sample is still limited: 
The late-time nebular spectra are available for 
$\sim 20$ SNe Ia in the optical wavelength. There are only a few examples in 
the NIR and Mid-IR. For most of them, a temporal sequence of late-time spectra 
are not available. In this section, we highlight the importance of expanding 
the sample of late-time nebular spectra of SNe Ia. 

In doing this, we summarize predictions from recent theoretical models. 
Two scenarios have been proposed to create the ECAP zone in the context of 
the asymmetric ignition of the initial deflagration bubbles. In the first scenario, 
many ignition points are clustered near the center of the white-dwarf with possible 
bulk off-set (\S 1), producing the ECAP zone by the deflagration flame. 
In the second scenario, one or a few ignition points are placed far away from 
the center of the white dwarf. The deflagration is weak and does not produce much 
neutron-rich materials. The deflagration flame rises to the surface without initiating 
the detonation.  The detonation may then be triggered near the white dwarf 
surface, at the opposite side of the initial deflagration ignition (so-called Gravitationally 
Confined Detonation model; Jordan et al. 2008). In this case, 
the initial deflagration is so weak and the white dwarf suffers from 
virtually no expansion before the detonation passes through the central region. 
Thus, the electron capture reactions in the detonation are not negligible, producing 
the ECAP zone (Meakin et al. 2009). This model predicts the off-set of the 
ECAP zone toward the opposite direction of the initial deflagration ignition. 
This second scenario could actually be regarded as an extreme case of the 
first scenario in terms of the distribution of the initial deflagration bubbles. 

Although these two scenarios are qualitatively different to one another for the origin of 
the ECAP zone, the resulting configuration 
(e.g., the off-center ECAP zone) may be similar to what we derived in this paper. 
In the second scenario, however, the detonation should produce a large amount of $^{56}$Ni 
in order to produce the ECAP zone more massive than $\sim 0.1 M_{\odot}$ 
(Meakin et al. 2009), at least their model sequence presented to date 
(e.g., $\sim 1.1 M_{\odot}$ of $^{56}$Ni in their 2D models). 
This scenario therefore may not be favored as the origin of 
the ECAP zone for the particular case of SN 2003hv, 
which is relatively faint as a normal SN Ia (Leloudas et al. 2009). 
This second scenario is still a possibility to explain the wavelength shift in the nebular 
emission lines seen in bright SNe Ia, which also show the signature of 
the asymmetry in the ECAP zone (\S 5 \& \S 6.1). 
Although we expect that the Gravitationally 
Confined Detonation (GCD) model results in bright SNe Ia if the detonation is 
to produce the ECAP zone, the detailed relation between the brightness and the mass of the 
ECAP zone should be dependent on details of the initial conditions such as 
the ignition density (which depends on the accretion rate of a white-dwarf before the 
explosion; e.g., Nomoto 1982; Nomoto et al. 1984). Further study based on the GCD scenario 
is therefore important. 

We suggest that future observations of late-time SN Ia emission can provide 
important information to 
understand the details of the explosion mechanism(s). Hereafter, we discuss 
importance of the following (future) observations, especially in view of the 
available theoretical ideas as mentioned above. 
\begin{itemize}
\item Expanding the sample of the line velocity shift measurement, 
especially in the optical wavelengths. 
\item Studying the detailed line profiles, especially in the NIR. 
\item Investigating the mass of the ECAP zone. 
\end{itemize}

\subsubsection{Expanding the sample for the line velocity shift measurement}

A series of hydrodynamic simulations based on the first scenario 
provide interesting predictions (Kasen et al. 2009). 
More asymmetric distribution of the initial deflagration 
results in a larger amount of $^{56}$Ni produced by the detonation, yielding brighter 
SNe. For a larger degree of the off-set in the initial deflagration bubbles, we should 
also expect a larger degree of the off-set in the ECAP zone produced by the deflagration 
(Maeda et al. 2009c). We therefore expect 
that brighter SNe Ia show a larger off-set velocity in the ECAP zone in this scenario.  
Since the wavelength shift seen in nebular spectra of individual SNe Ia is 
a result from a combination of the degree of the asymmetry and the viewing angle, 
this relation should be observationally investigated in a statistic way with a large number 
of SNe Ia. 
We suggest to investigate the distribution of the line-velocity (Figure 10) as a function 
of the luminosity of SNe Ia (or so-called $\Delta m_{15}$ in the light curve shape which 
correlates with the luminosity). The scenario predicts that brighter SNe Ia 
should show a wider distribution in the velocity space, while fainter SNe Ia should show 
more narrowly peaked distribution at zero velocity. 

The GCD model is an extreme end of the model 
sequence, predicting bright SNe Ia. 
The mechanism to create the ECAP zone is different from the first scenario. 
Thus, it may well give 
the distribution of the velocity shift as a function of the luminosity different from 
the first scenario. 

These issues cannot be addressed with the current sample of SNe Ia for which nebular spectra 
are available -- the current data are consistent with the idea that SNe Ia have 
the generic off-set as derived for SN 2003hv (Fig. 10), but it does not reject a possibility 
that the off-set velocity is dependent on the luminosity because of the small sample. 
Increasing the number of the sample will tell us (1) whether brighter 
SNe Ia have larger off-set in the ECAP zone (a test for the first explosion scenario), 
and (2) whether the distribution of off-set of the ECAP zone is explained by a single 
scenario or hybrid scenarios, especially in bright SNe Ia (a test 
for the GCD model). 

\subsubsection{Detailed Line Profiles especially in the NIR}

In this paper, we have focused on the line wavelength shift, but have not 
tried to fit the details of the line profiles. 
H\"oflich et al. (2004) and Motohara et al. (2006) suggested to use 
the detailed line profile of [Fe~II]~$1.644~\micron$ to 
probe the distribution of 
electron capture products (i.e., the ECAP-zone in our model). 
H\"oflich et al. (2004) 
suggested that the apparently flat-topped profile of this line 
in SN 2003du is a signature of a 
"hole'' in the region emitting the [Fe~II]. They attributed this hole 
to the existence of an inner $^{56}$Ni-free, electron capture products-rich 
region, since such a region lacks the heating source when the ejecta become
transparent to $\gamma$-rays. 
The same suggestion was also made for SN 2003hv based on better data 
(Motohara et al. 2006). 
There are also other possibilities for the flat-topped profile, e.g., 
the infrared catastrophe (Leloudas et al. 2009). 

Understanding the origin of the details of the line profiles requires further studies 
(e.g., H\"oflich et al. 2006). 
For the "hole" interpretation of the flat-topped profiles, 
it should be examined whether such a configuration can result from hydrodynamic 
explosion models. 

An intensive model survey based on the first scenario has been presented 
by Kasen et al. (2009) in 2D. According to their model sequence, we would expect that 
more asymmetric distribution of the initial deflagration bubbles could result in 
a larger amount of $^{56}$Ni produced by the detonation near the zero-velocity. 
As such, we expect that bright SNe Ia tend to show the peaked line profiles. 
Faint SNe Ia may tend to show the flat-topped profile (see below). 
The most solid case showing the flat-top profile in the NIR [Fe~II] so far is SN 2003hv. 
This is a relatively faint SN Ia, thus being consistent with this theoretical expectation. 
Also, the bright SN 1991T showed the peaked profile (Bowers et al. 1997) and it is 
also consistent with the expectation. 

Apart from the above prediction, 
an issue remains whether the first scenario can produce the flat-topped profiles. 
The current state-of-art 3D deflagration model according to this scenario 
indicates some amount of 
unburned C+O is mixed down to the center (e.g., R\"opke et al. 2007a). 
These materials may be later detonated to produce $^{56}$Ni 
(R\"opke \& Niemeyer 2007b; Kasen et al. 2009). 
Thus, it is expected that the ECAP zone and the HD zone are mixed, 
rather than form the distinctly layered structure. 
An interesting question is whether future simulations can lead to the "hole" configuration 
with some initial conditions for the thermonuclear bubbles, 
and whether different initial conditions can result in a different degree of the mixing 
(to explain the diversity in the detailed profile). 

In the second, GCD scenario, the ECAP zone would not suffer from the mixing process since 
it is created by the detonation. It is expected to be always a "hole", but the size of the zone 
may well depend on the strength of the initial deflagration. For the weaker deflagration, 
the density of the white-dwarf at the detonation 
is higher, and thus the size and mass of the ECAP zone are larger 
(Meakin et al. 2009). 
This raises an interesting possibility that the variation in the size of the ECAP zone 
may explain the variation in the detailed line profile; brighter SNe Ia are expected to 
more likely produce a flat-top profile. 
This tendency is different from the expectation from the first scenario.

In our kinematic 
model, we have the "hole" in the form of the ECAP-zone, which is assumed
to be macroscopically separated from the outer $^{56}$Ni-rich
HD/LD-zones. In this configuration,  our model makes the following predictions;  
(1) Emission lines formed in the ECAP-zone have
narrow widths ($\sim 3,000$~km~s$^{-1}$),  and have a peaked
profile (with respect to the "off-set" wavelength). 
This is exemplified by [Ni~II]~$\lambda$7378, [Ni~II]~$6.634~\micron$, 
[Ni~III]~$7.350~\micron$.  
(2) Emission lines originating from the HD-zone
have flat-topped profiles.  
Typical examples are [Fe~II]~$\lambda$7151, [Fe~II]~$1.257~\micron$, 
[Fe~II]~$1.644~\micron$, and [Co~III]~$11.88~\micron$.  
(3) Emission lines dominated by the LD-zone have 
broad, peaked profiles, since the size of the "hole" is much smaller
than the emitting radius.  Typical examples are 
[Fe~III]~$2.218~\micron$, and [Fe~III]~$2.904~\micron$. 
The [Fe~III] blend at 4700\AA\ also has the same tendency, 
although the blended nature of this feature makes the situation complicated. 
Some lines show a combination of these characteristics; for example, 
[Fe~II]~$17.93~\micron$ is predicted to show a peaked, relatively broad profile
with a  wavelength shift depending on the viewing angle (\S
5.4). Thus, looking at various emission lines, we will be able to obtain
different, independent information on the  explosion physics.

The best lines for investigating the details of the distribution of $^{56}$Ni 
and the hole (i.e., the ECAP zone) 
are [Fe~II]~1.257~$\micron$ and [Fe~II]~1.644~$\micron$, thanks to 
their low excitation temperature and the relatively isolated nature of these lines. 
Adding to this, [Ni~II]~$\lambda$7378 and some Mid-IR lines (Table 4) 
can give the direct view of the ECAP zone distribution. 
Thus, expanding the sample of late-time NIR spectra is highly encouraged, 
preferentially with the optical and (if possible) mid-IR data taken almost simultaneously. 
This will tell us (1) whether brighter 
SNe Ia have a larger amount of the low-velocity $^{56}$Ni (a test for the first explosion scenario), 
(2) whether the fainter SNe tend to show the flat-topped profile  
(a constraint on the distribution of the deflagration bubbles),  
and (3) whether the brightest SNe Ia show the hole of the $^{56}$Ni distribution 
(a test for the GCD model as the origin of the brightest SNe Ia).

\subsubsection{Mass of Neutron-rich Fe-peak Elements} 

In \S 6.2, we have mentioned an importance of multi-epoch spectroscopy 
to clarify details of thermal processes within SN Ia nebulae. 
This is also important to 
clarify the explosion physics. 

Different scenario should result in the different mass of 
the ECAP zone. The first scenario predicts a smaller mass of the ECAP zone 
for brighter SNe Ia (since the weaker detonation is followed by the stronger detonation; 
Kasen et al. 2009). The trend should be opposite in the GCD model sequence 
(since the neutron-rich materials and $^{56}$Ni are both created by the detonation). 
Deriving the mass of neutron-rich materials (i.e., ECAP zone) as a function 
of the SN Ia peak luminosity will therefore give us a strong test for these scenarios. 
No significant variation of the mass of neutron-rich materials as a function of 
the luminosity (or $\Delta m_{15}$) has been found so far (Mazzali et al. 2007). 
Understanding the thermal properties will provide the better accuracy 
in the mass estimate so that the investigation of the details of the explosion physics 
becomes possible. 

\acknowledgements

The authors would like to thank Friedrich K. R\"opke and 
Wolfgang Hillebrandt for useful discussion. 
The authors are grateful to Christopher L. Gerardy for the {\it Spitzer} data. 
This research is supported by World Premier International Research Center
Initiative (WPI Initiative), MEXT, Japan. 
K.M. acknowledges financial support by Grant-in-Aid for
Scientific Research for young Scientists (20840007) and by the Max-Planck Society 
as a short-term visitor. 
S.T. acknowledges support by the Transregional Collaborative Research Centre TRR 33 `The Dark Universe' of the German Research Foundation (DFG).
J.S. is a Royal Swedish Academy of Sciences Research Fellow supported by a grant from the Knut and Alice Wallenberg Foundation. The Dark Cosmology Centre is funded by the Danish 
National Research Foundation.
The work has also been supported by the Grant-in-Aid for
Scientific Research of the JSPS (20540226) and MEXT (19047004, 20040004). 
This research made use of the {\it SUSPECT} (the online Supernova Spectrum Archive), 
maintained at the Department of Physics and Astronomy, University of Oklahoma. 
The authors acknowledge the CfA Supernova Archive, which is funded in part by the National Science Foundation through grant AST 0606772, for the data of SN 2001V.

\end{document}